\documentclass{webofc}
\usepackage[varg]{txfonts}   
\begin{document}
\title{Measurement of $\mathbf{\phi_{s}}$ and $\mathbf{\Delta\Gamma_{s}}$ at LHCb}
%
%

\author{\firstname{Varvara} \lastname{Batozskaya}\inst{1}\fnsep\thanks{\email{varvara.batozskaya@cern.ch}} 
\\ \scriptsize{on behalf of the LHCb Collaboration}
}

\institute{National Centre for Nuclear Research (NCBJ), Warsaw, Poland}

\abstract{%
Determination of the mixing-induced $C\!P$-violating phase $\phi_{s}$ and decay width difference $\Delta\Gamma_{s}$ in $\bar{b}\to\bar{c}c\bar{s}$ decays is one of the main goals of the LHCb experiment. Thanks to the precise prediction of the $\phi_{s}$ value within the Standard Model, it represents an excellent probe to search for new physics. The measurements of $\phi_{s}$ and $\Delta\Gamma_{s}$ at LHCb are reviewed including results from the 3.0~fb$^{-1}$ dataset accumulated during 2011-2012. Further measurement improvement is expected from the inclusion of results obtained using decay modes with smaller branching fraction.}
\maketitle
\section{Introduction}
\label{intro}
\begin{figure}[htb]
\centering
\begin{minipage}{0.24\linewidth}
  \centerline{\includegraphics[width=92pt]{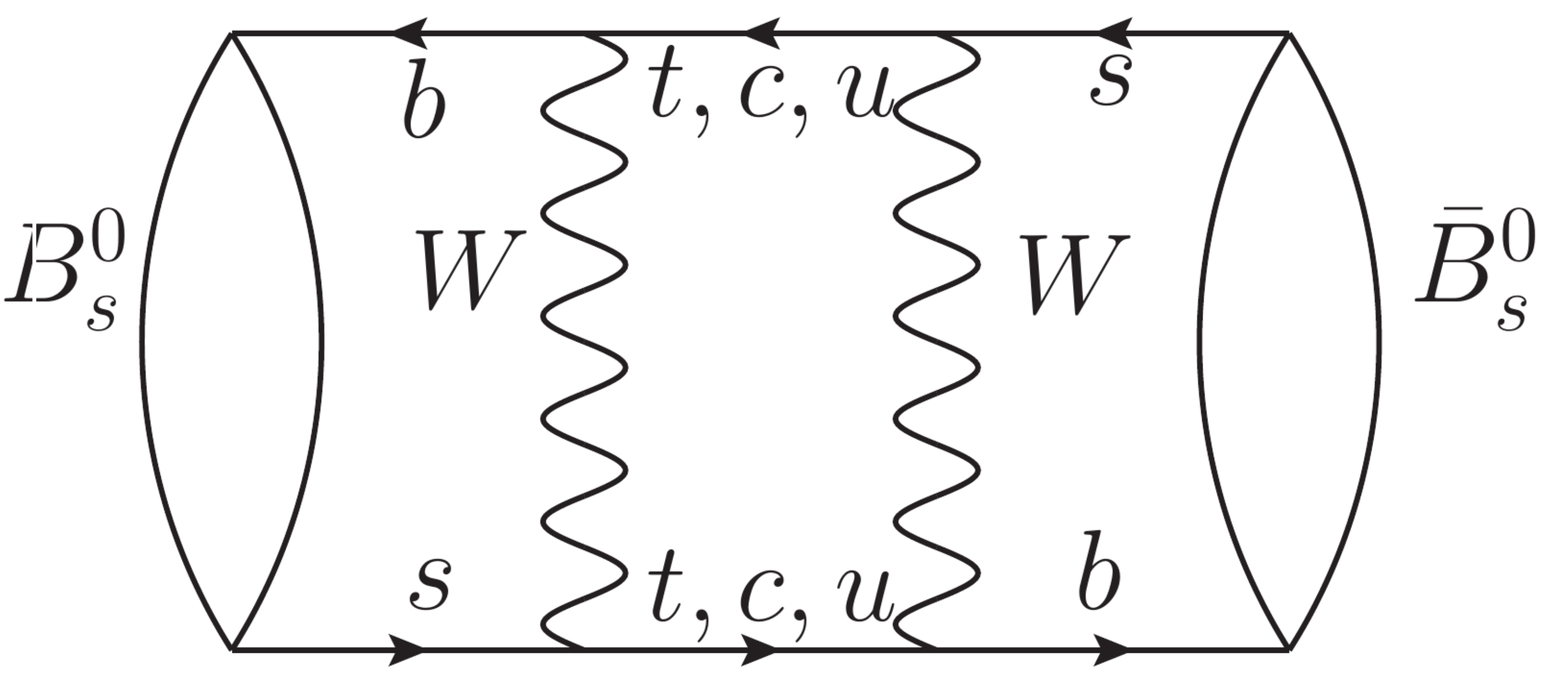}\put(-55,-25){(a)}}
  \end{minipage}
 \hfill
   \begin{minipage}{0.24\linewidth}
  \centerline{\includegraphics[width=91pt]{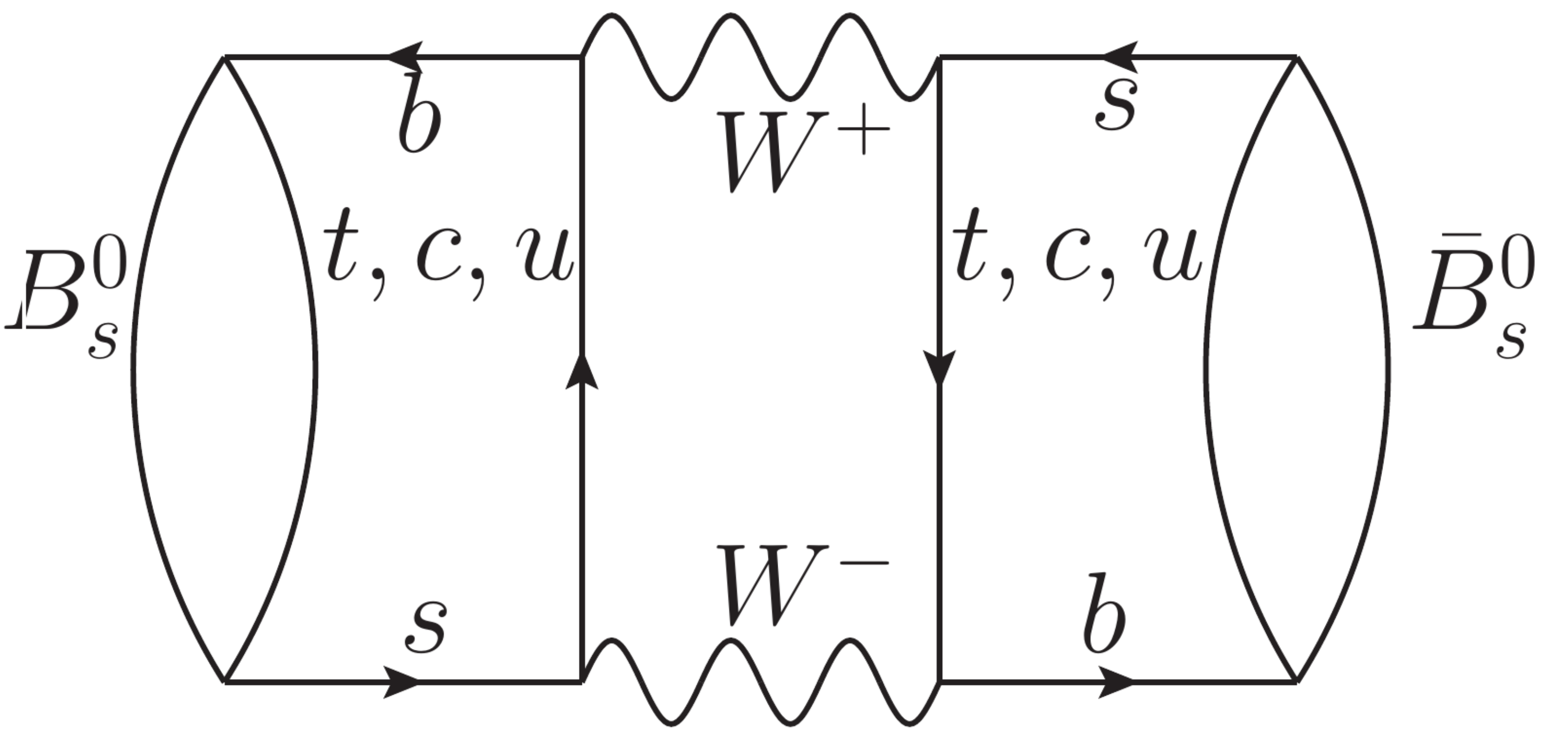}\put(-55,-22){(b)}}
 \end{minipage}
 \hfill
   \begin{minipage}{0.24\linewidth}
  \centerline{\includegraphics[width=92pt]{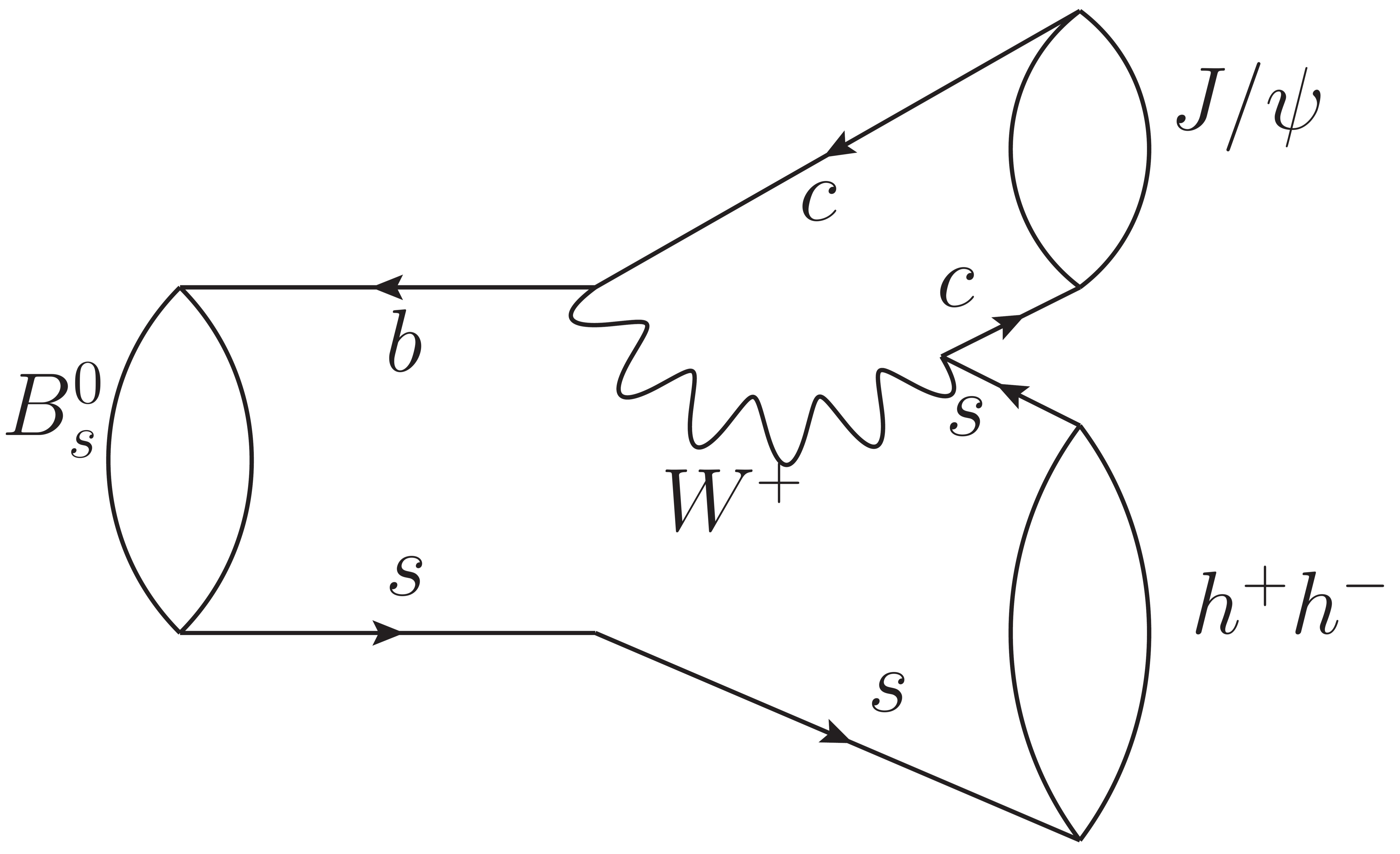}\put(-55,-9){(c)}}
   \end{minipage}
 \hfill
   \begin{minipage}{0.24\linewidth}
  \centerline{\includegraphics[width=92pt]{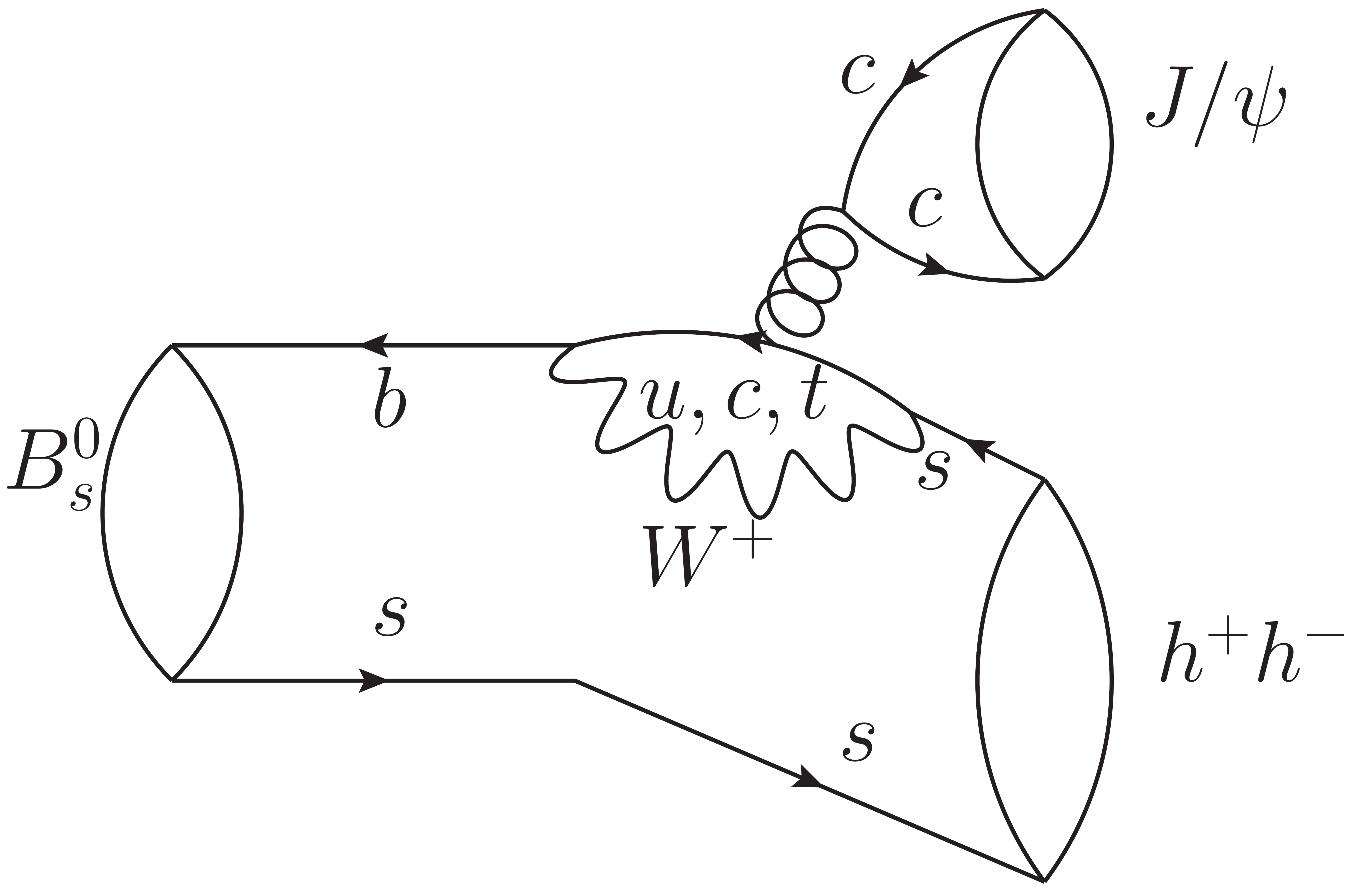}\put(-55,-4){(d)}}
 \end{minipage}
\caption{Feynman diagrams: (a-b) $B^{0}_{s}-\bar{B}^{0}_{s}$ mixing and contributions to the decay $B^{0}_{s}\to J/\psi h^{+}h^{-}$ within the SM, where $h = \pi, K$: (c) "tree-level" and (d) "penguin" diagrams.}
\label{Fig:Mixing}
\end{figure}

The $C\!P$-violating phase $\phi_{s}$ can be related to the angle $\beta_{s}$ of the unitary Cabbibo-Kobayashi-Maskawa (CKM) triangle of the $B^{0}_{s}$ meson system analogous to $\beta$ angle in $B^{0}$ meson decay~\cite{Cabibbo:1963yz}. The interference between the direct decay of $B^{0}_{s}$ mesons to $C\!P$ eigenstates via $\bar{b}\to \bar{c}c\bar{s}$ transitions and $B^{0}_{s}-\bar{B}^{0}_{s}$ mixing allows to measure the phase $\phi_{s}$: 

\begin{equation}
 \phi_{s} = \phi_{M}-2\phi_{D} = -2\beta_{s} + \Delta\phi^{Peng}_{s} + \delta^{NP}_{s}
\end{equation}
where $\phi_{M}$ and $\phi_{D}$ are the mixing and direct phases, respectively. The value of $\phi_{s}$ can be shifted with respect to the Standard Model (SM) value by the presence of higher order "penguin" diagrams from non-perturbative hadronic effects (Fig.~\ref{Fig:Mixing}) and new physics (NP) contributions that could be difficult to distinguish from "penguins". These components start to play an important role when reaching high precision of the $\phi_{s}$ measurement~\cite{bib:Ligeti}.  

If only the dominant "tree-level" contributions are included (Fig.~\ref{Fig:Mixing}), the phase $\phi_{s}$ within the Standard Model is predicted to be $-2\beta_{s}$ where $\beta_{s}=\arg(-V_{ts}V^{\ast}_{tb}/V_{cs}V^{\ast}_{cb})$~\cite{bib:Kobayashi}. An indirect determination of $\phi_{s} = -37.6^{+0.7}_{-0.8}$~mrad is obtained using a global fit to experimental data~\cite{Charles:2015gya}.

\section{Status of $\phi_{s}$ and $\Delta\Gamma_{s}$ measurement}
\label{sec-1}
\subsection{$B^{0}_{s}\to J/\psi\phi$ analysis and combination with $B^{0}_{s}\to J/\psi\pi^{+}\pi^{-}$}

The phase $\phi_{s}$ and the decay width difference $\Delta\Gamma_{s}$ are extracted using a tagged time-dependent angular fit to $B^{0}_{s}\to J/\psi(\to\mu^{+}\mu^{-})\phi(\to K^{+}K^{-})$ candidates as described in Ref.~\cite{bib:Aaij1}. The final state is decomposed into four polarization amplitudes: three P-waves, $A_{0}$, $A_{\parallel}$, $A_{\perp}$ and one S-wave, $A_{S}$ accounting for the non-resonant $K^{+}K^{-}$ configuration. The angular analysis is required to disentangle the interfering $C\!P$-even and $C\!P$-odd components in the final state which arise due to total spin conservation between two vector resonances coming from a pseudoscalar meson decay.  

 A sample of 95~690~$\pm$~350 signal $B^{0}_{s}\to J/\psi\phi$ candidates are obtained after the trigger and off-line selection. The fit procedure takes into account angular and decay time acceptances, decay time resolution as well as the tagging efficiency. A simulated sample is used to determine the angular acceptance. The decay time acceptance is defined from data, using a prescaled unbiased trigger sample and a tag-and-probe technique. The decay time resolution is estimated to be $\sim$45~fs using prompt $J/\psi K^{+}K^{-}$ combinations. The flavour tagging algorithms use the information from additional same-side and opposite-side particles with respect to the signal candidates optimised on simulated samples and calibrated on data using flavour specific control channels. The obtained effective tagging power is (3.73$\pm$0.15)$\%$~\cite{bib:Aaij1}.

\begin{figure}[htb]
\begin{minipage}{0.24\linewidth}
  \centerline{\includegraphics[width=106pt]{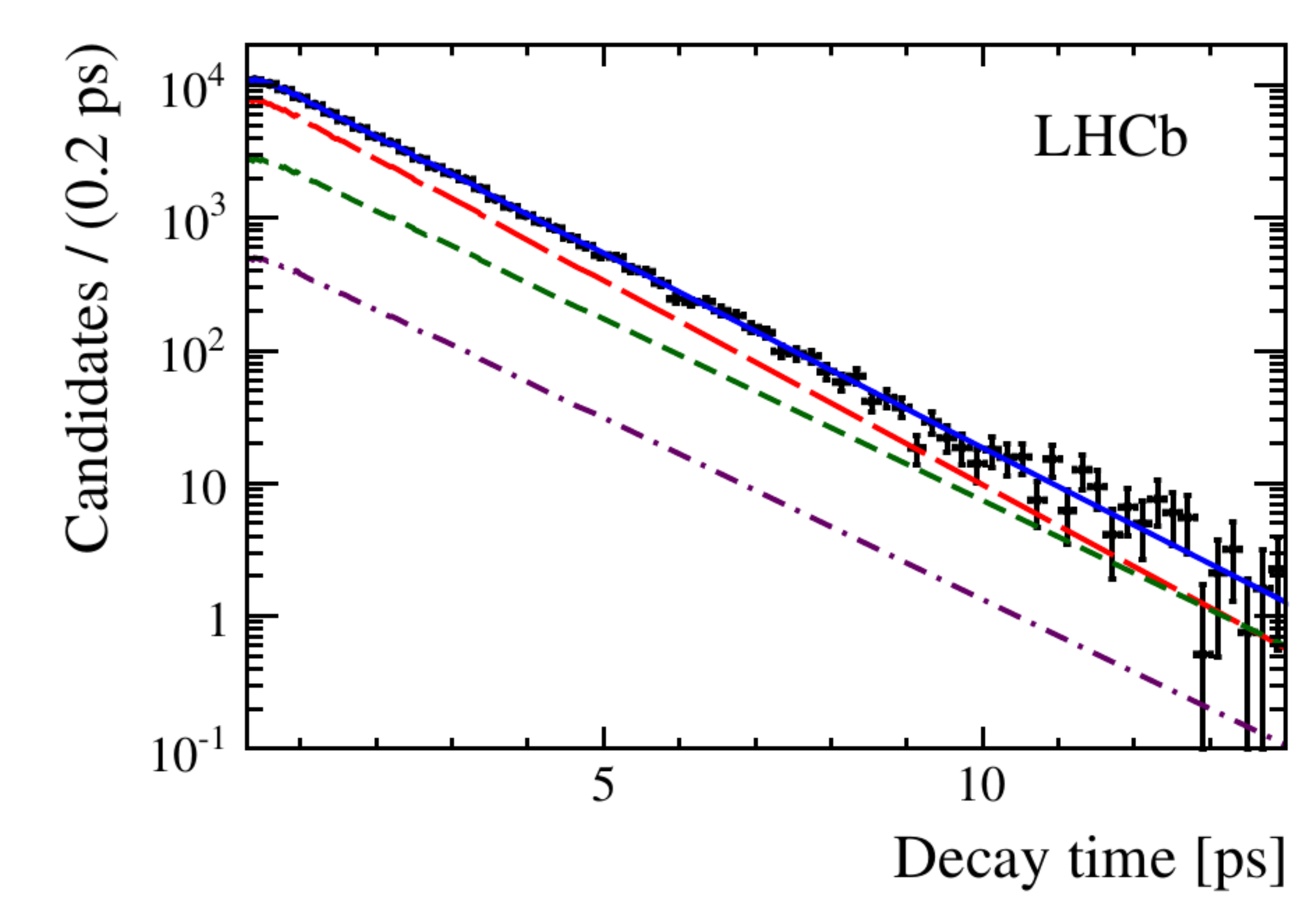}}
 \end{minipage}
 \hfill
   \begin{minipage}{0.24\linewidth}
  \centerline{\includegraphics[width=100pt]{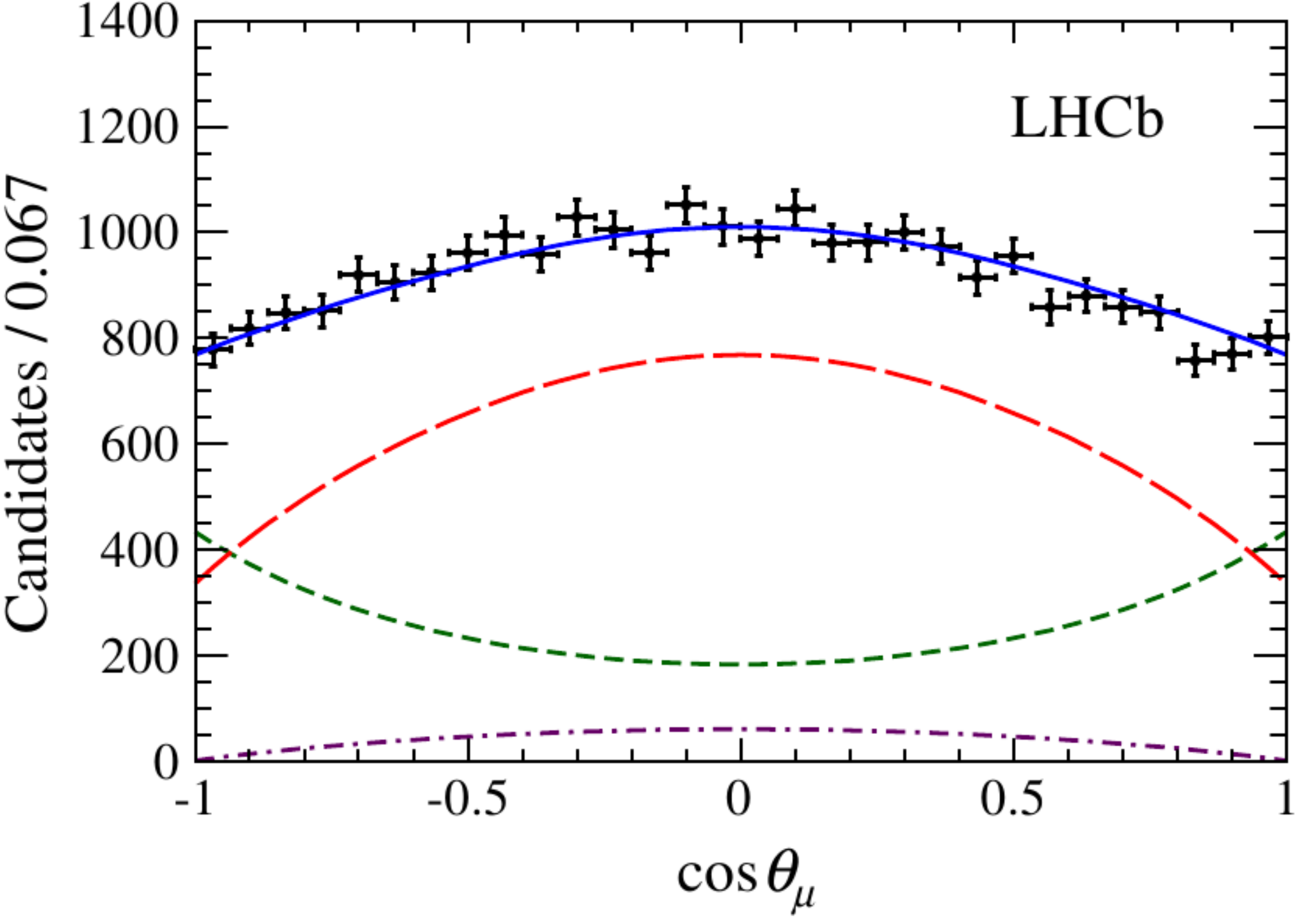}}
 \end{minipage}
 \hfill
 \begin{minipage}{0.24\linewidth}
  \centerline{\includegraphics[width=100pt]{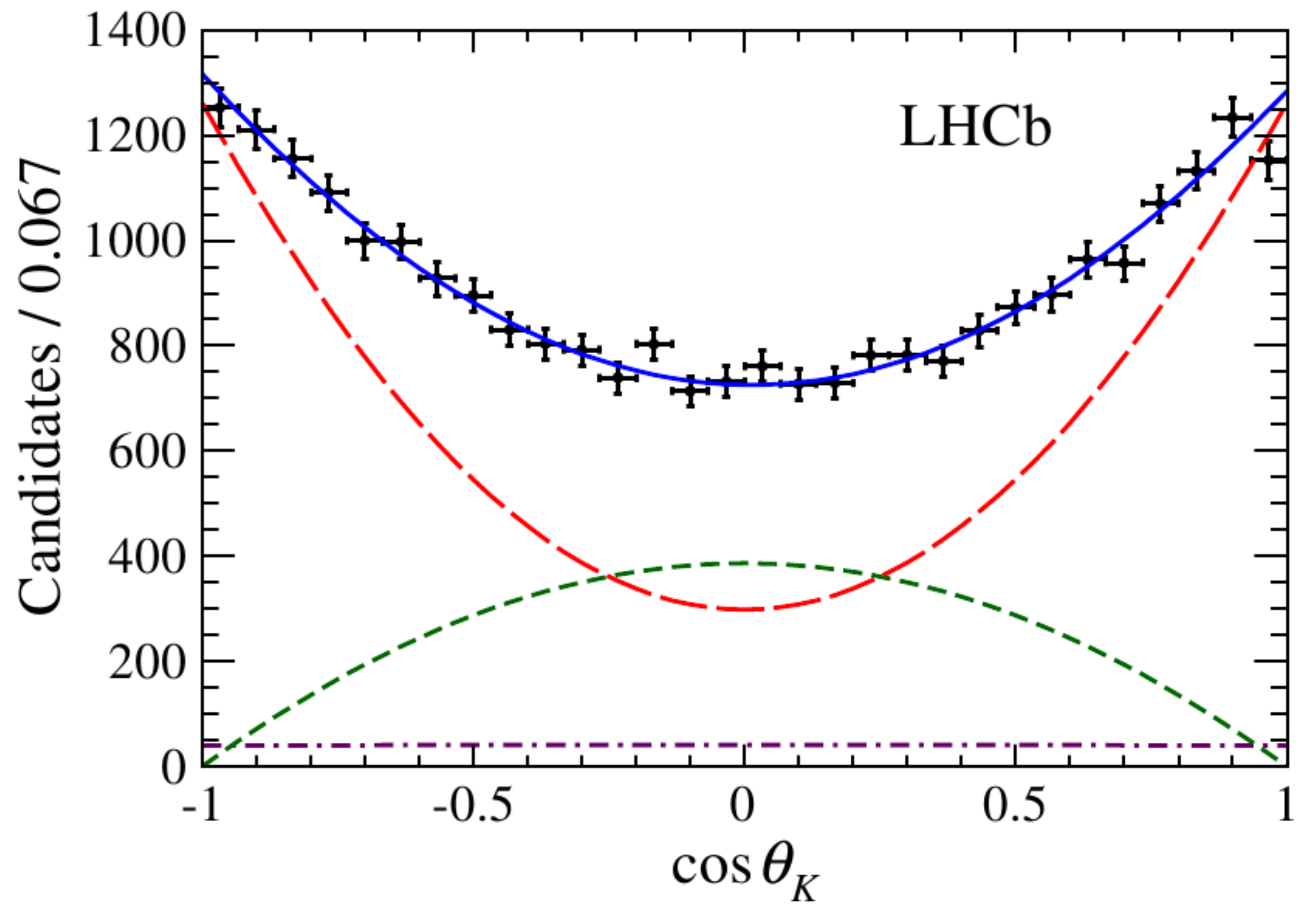}}
 \end{minipage}
 \hfill
   \begin{minipage}{0.24\linewidth}
  \centerline{\includegraphics[width=100pt]{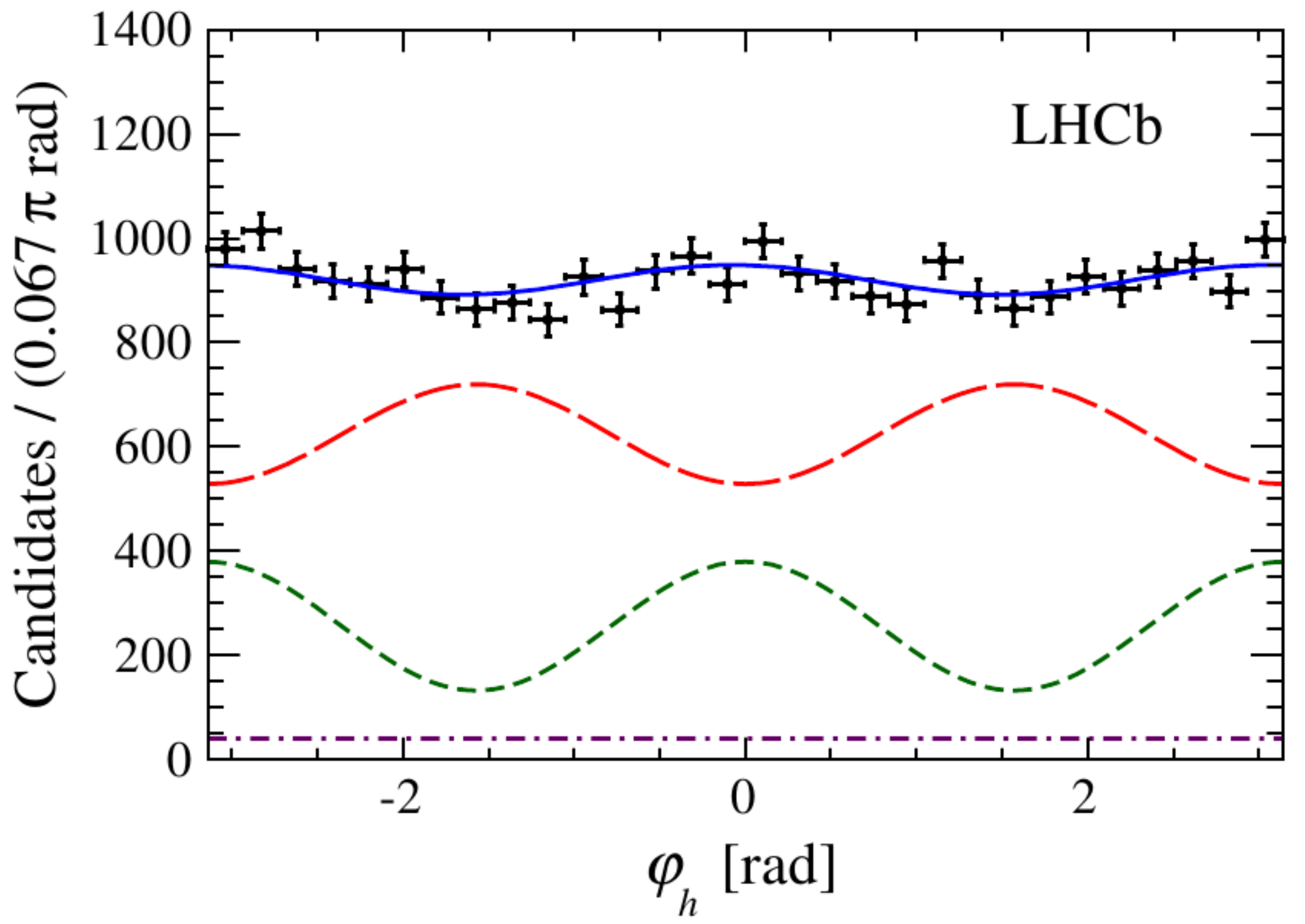}}
 \end{minipage}
 \caption{Decay time and angle distributions for $B^{0}_{s}\to J/\psi\phi$ decays (black markers) with the one-dimensional projections of the PDF. The solid blue line shows the total signal contribution, which is composed of $C\!P$-even (long-dashed red), $C\!P$-odd (short-dashed green) and S-wave (dotted-dashed purple) contributions.}
\label{Fig:TimeAngles}
\end{figure}

A weighted unbinned likelihood fit is performed using a signal-only PDF as described in Ref.~\cite{bib:Xie}. The signal weights are extracted using the sPlot technique~\cite{bib:Pivk}. The fit is divided into six bins of $m(K^{+}K^{-})$ region to allow the measurement of the small ($\sim$2$\%$) S-wave amplitude in each bin and to minimize correction factors due to the interference between the different components of the final state. The projections of the decay time and angular distributions are shown in Fig.~\ref{Fig:TimeAngles}. The measured results are $\phi_{s}=-58\pm49\pm6$~mrad and $\Delta\Gamma_{s}=0.0805\pm0.0091\pm0.0032$~ps$^{-1}$, where the first uncertainty is statistical and the second systematic~\cite{bib:Aaij1}. This measurement of the $C\!P$-violating parameter, $\phi_{s}$, is the single most precise to date and is in agreement with the SM predictions~\cite{Charles:2015gya, Artuso:2015swg}. The dominant source of systematic uncertainty comes from knowledge of the angular and decay time efficiencies, respectively. 

The $B^{0}_{s}\to J/\psi\pi^{+}\pi^{-}$ decay analysis~\cite{bib:Aaij3} is similar to the $B^{0}_{s}\to J/\psi\phi$ one with a noticeable simplification. The angular analysis is not needed because the final state has been found to be $>97.7\%$ completely $C\!P$-odd with $f_{0}(980)$ representing the dominant component~\cite{LHCb:2012ae}. A combination of the $B^{0}_{s}\to J/\psi\phi$ and $B^{0}_{s}\rightarrow J/\psi\pi^{+}\pi^{-}$ measurement gives the result of $\phi_{s}=-10\pm39$~mrad~\cite{bib:Aaij1}. 

\subsection{$B^{0}_{s}\to\psi(2S)\phi$ analysis}

\begin{figure}[htb]
\begin{minipage}{0.48\linewidth}
  \centerline{\includegraphics[width=130pt]{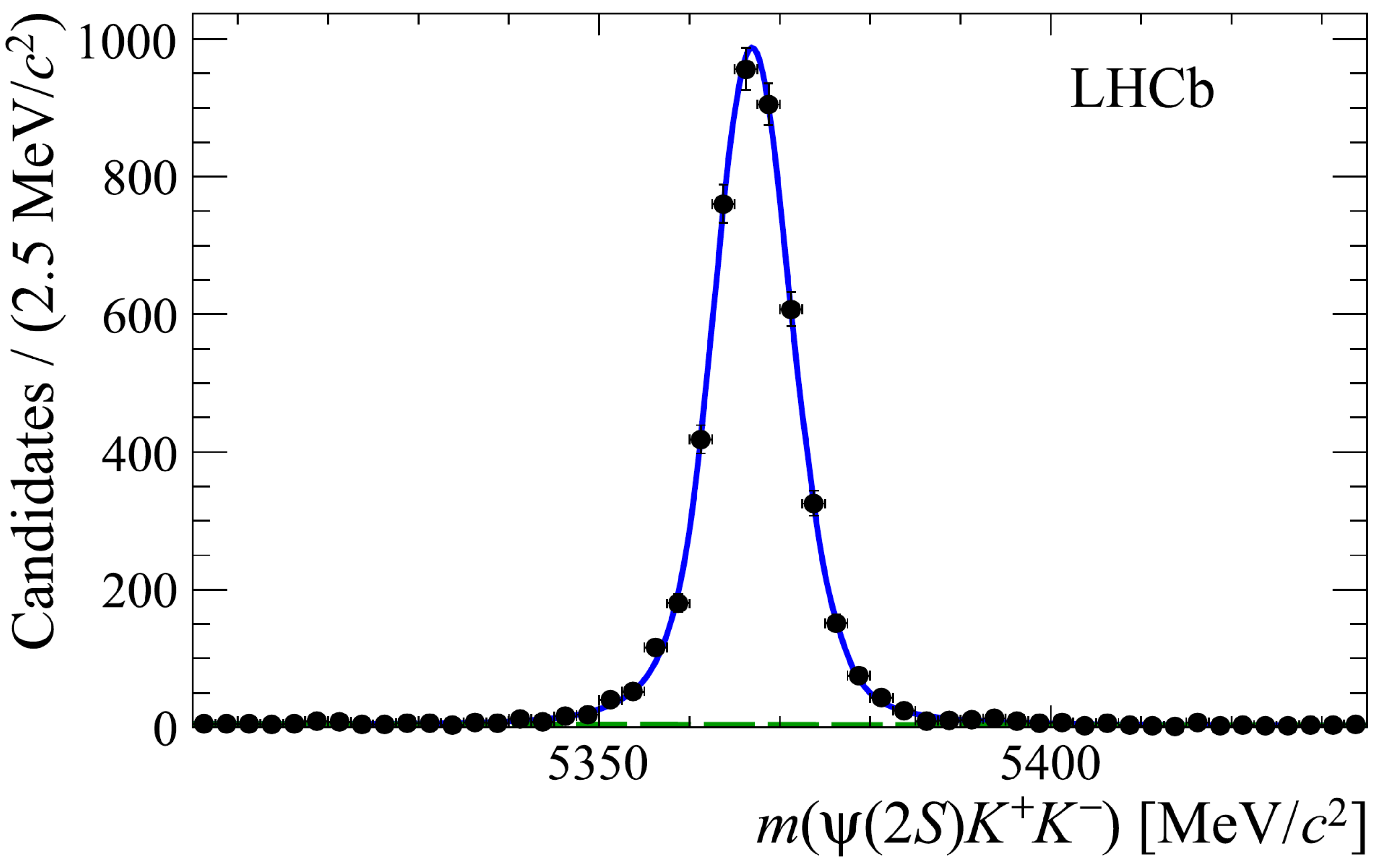}}
 \end{minipage}
 \hfill
   \begin{minipage}{0.48\linewidth}
  \centerline{\includegraphics[width=130pt]{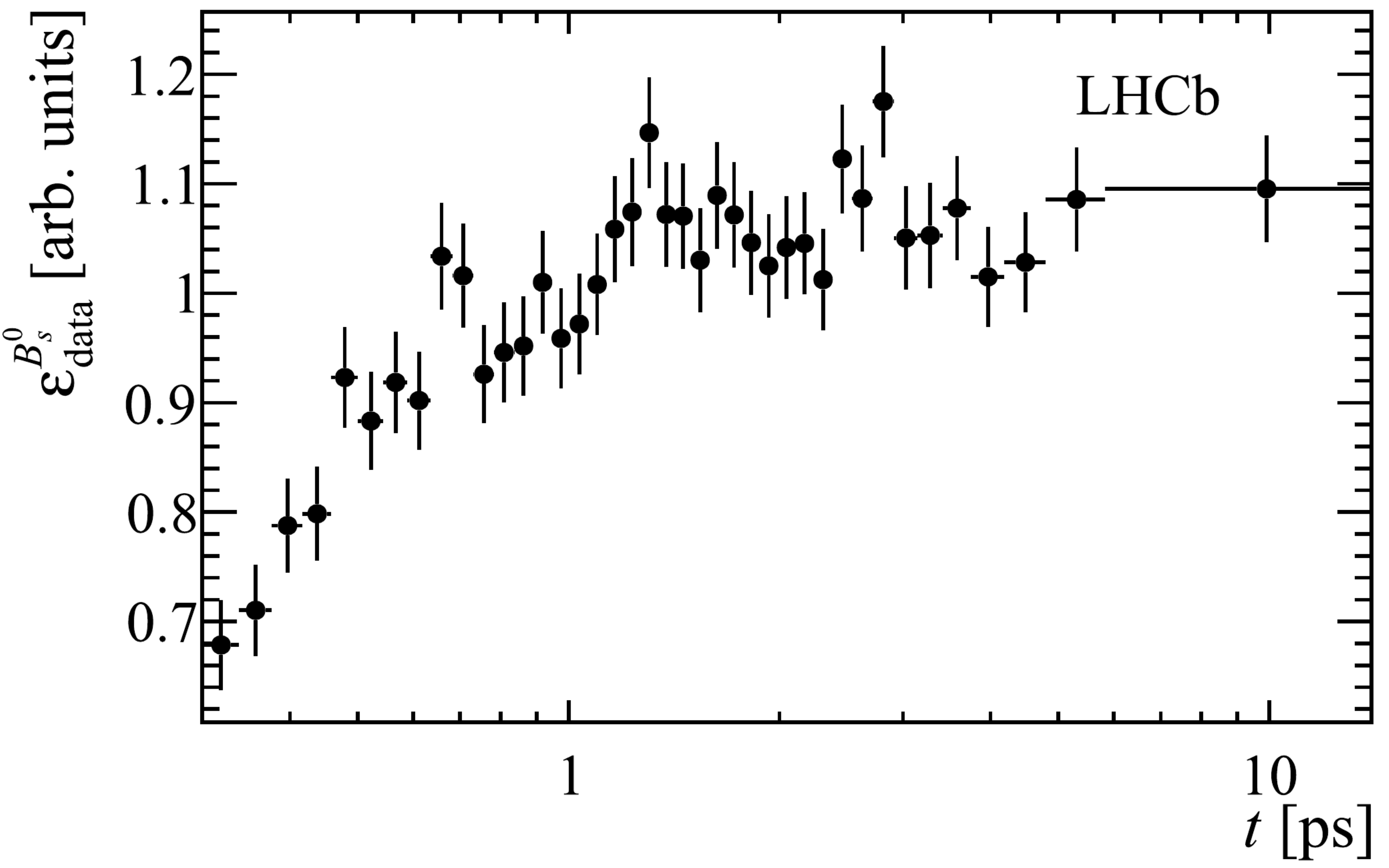}}
 \end{minipage}
\caption{Distribution of the $m(\psi(2S)K^{+}K^{-})$ invariant mass for the selected $B^{0}_{s}\to\psi(2S)\phi$ candidates and decay time acceptance in arbitrary units.}
\label{Fig:Psi2SPhi}
\end{figure}

The  $B^{0}_{s}\to\psi(2S)(\to\mu^{+}\mu^{-})\phi(\to K^{+}K^{-})$ is another decay mode with $\bar{b}\to\bar{c}c\bar{s}$ transition that has been exploited by the LHCb collaboration to measure $\phi_{s}$ and $\Delta\Gamma_{s}$~\cite{Aaij:2016ohx}. The formalism used for this analysis is very close to that of $B^{0}_{s}\to J/\psi\phi$ decay~\cite{bib:Aaij1} where the $J/\psi$ meson is replaced with $\psi(2S)$. The number of signal candidates selected from a fit to the data sample is $\sim4700$ (Fig.~\ref{Fig:Psi2SPhi}). The decay time acceptance is determined using a control $B^{0}\to\psi(2S)K^{\ast0}(\to K^{+}\pi^{-})$ decay mode as shown in Fig.~\ref{Fig:Psi2SPhi}. The first measurement of the $C\!P$-violating parameters in a final state containing the $\psi(2S)$ resonance is $\phi_{s}=-230^{+290}_{-280}\pm20$~mrad and $\Delta\Gamma_{s}=0.066^{+0.041}_{-0.044}\pm0.007$~ps$^{-1}$. The fit result is consistent with $B^{0}_{s}\to J/\psi\phi$ measurement and the SM predictions. The systematic uncertainty is less than 20$\%$ of the statistical uncertainty.
   
\subsection{$B^{0}_{s}\to J/\psi K^{+}K^{-}$ in high $m(K^{+}K^{-})$ range}

\begin{figure}[htb]
  \centerline{\includegraphics[width=140pt]{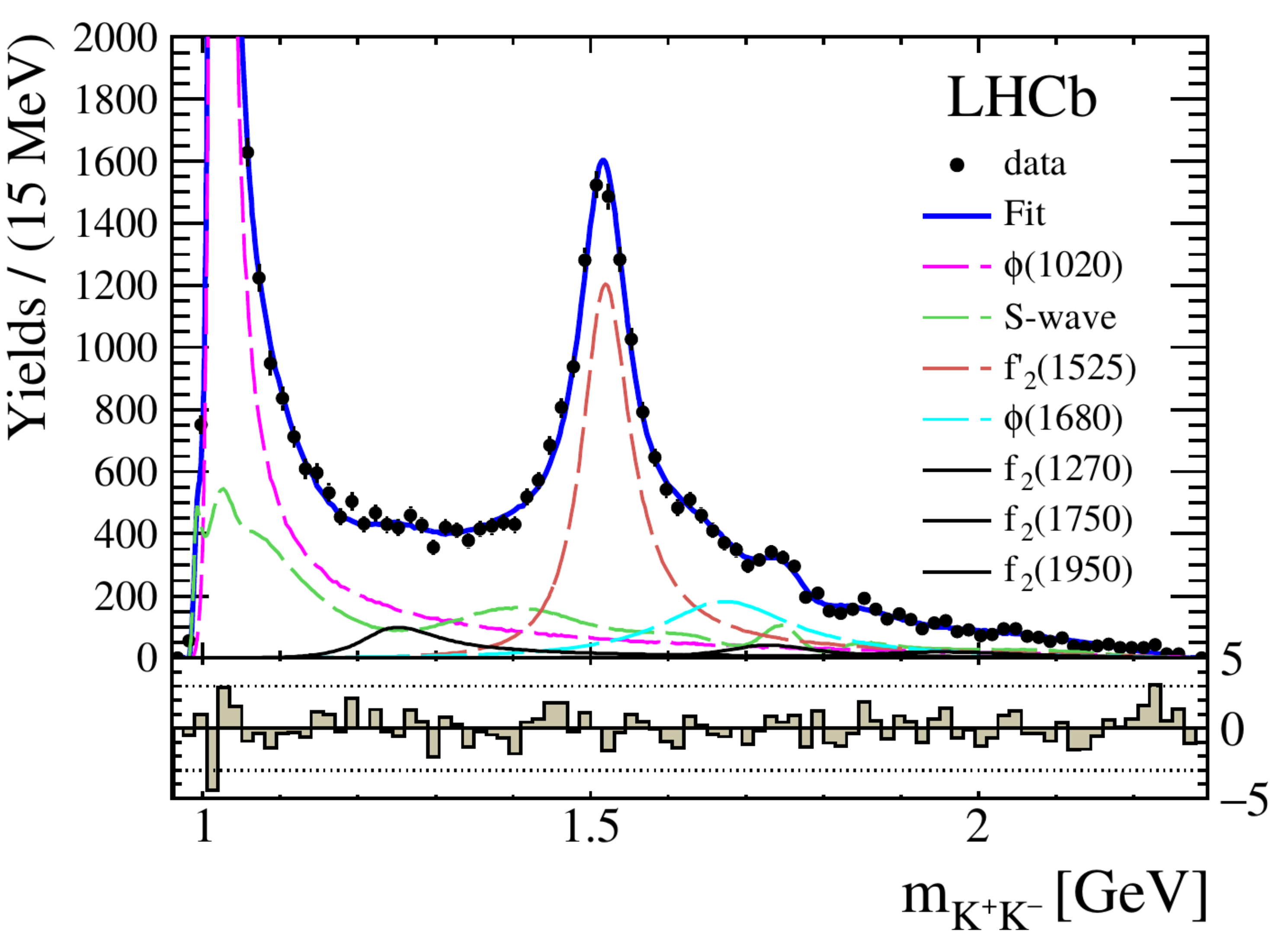}}
\caption{Distribution of the $m(J/\psi K^{+}K^{-})$ invariant mass with contributing components.}
\label{Fig:JpsiKKHM}
\end{figure}

The measurement of the $C\!P$-violating parameters has been also performed in the $B^{0}_{s}\to J/\psi K^{+}K^{-}$ decay with $K^{+}K^{-}$ invariant mass higher than 1050~MeV/c$^{2}$~\cite{Aaij:2017zgz} that is above the $\phi(1020)$ resonance region. The important difference between both decay analyses is that modelling of the $m(K^{+}K^{-})$ distribution is included to distinguish different resonant and nonresonant contributions. The decay time acceptance is determined with the same method as described in~\cite{Aaij:2016ohx} by using a control channel $B^{0}\to J/\psi K^{\ast0}(\to K^{+}\pi^{-})$. The $K^{+}K^{-}$ mass spectrum is fitted by considering the different contributions found in the time-dependent amplitude analysis as shown in Fig.~\ref{Fig:JpsiKKHM}. The final fit has been performed allowing eight independent sets of $C\!P$-violating parameters: three corresponding to $\phi(1020)$ transversity states, $K^{+}K^{-}$ S-wave, $f_{2}(1270)$, $f^{'}_{2}(1525)$, $\phi(1680)$ and the combination of the two high-mass $f_{2}(1750)$ and $f_{2}(1950)$ states. 
The measurement result of $B^{0}_{s}\to J/\psi K^{+}K^{-}$ in the high $m(K^{+}K^{-})$ region is $\phi_{s}=119\pm107\pm34$~mrad and $\Delta\Gamma_{s}=0.066\pm0.018\pm0.006$~ps$^{-1}$. The largest contribution to systematic uncertainty results from the resonance fit model. The combination with the $B^{0}_{s}$ decay fit results in the $\phi(1020)$ region gives $\phi_{s}=-25\pm45\pm8$~mrad that improves the precision of the $\phi_{s}$ measurement by more than 9$\%$.

\subsection{Global combination}

\begin{figure}[htb]
  \centerline{\includegraphics[width=160pt]{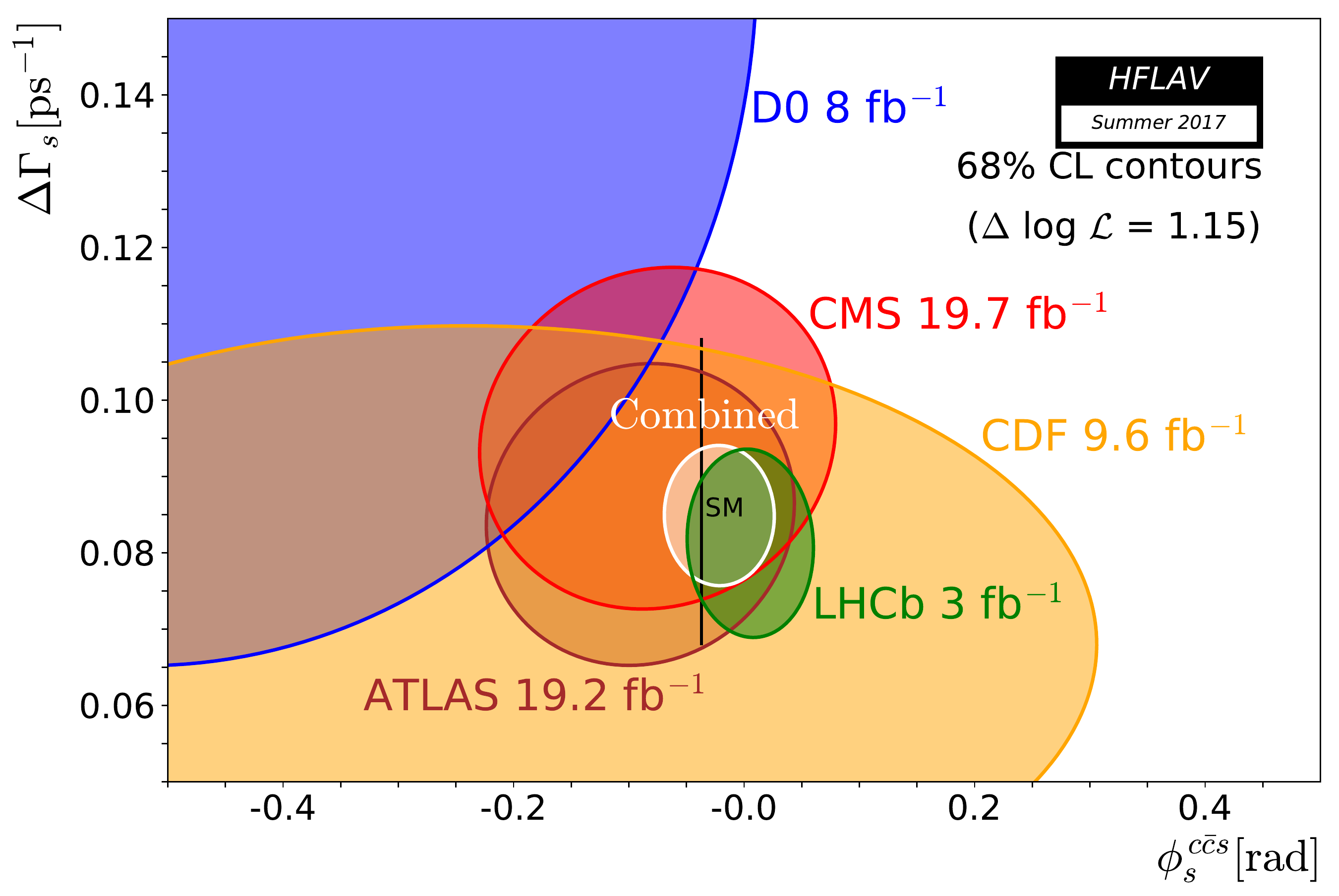}}
 \caption{68$\%$ confidence level regions in $\Delta\Gamma_{s}$ and $\phi_{s}$ plane obtained from individual contours of CDF, D0, CMS, ATLAS and LHCb measurements and the combined contour (solid line and shaded area)~\cite{bib:MainPlot}. The expectation within the SM~\cite{Charles:2015gya} is shown as a black thin rectangle.}
\label{Fig:MainPlot}
\end{figure}

The $C\!P$-violating phase and lifetime parameters have been measured by Tevatron and LHC experiments, namely four analysis using the $B^{0}_{s}\to J/\psi\phi$ final state from CDF~\cite{bib:CDF}, D0~\cite{bib:D0}, ATLAS~\cite{bib:ATLAS} and CMS~\cite{bib:CMS} collaborations and five analyses using different final states performed by the LHCb collaboration. The world average result of $\phi_{s}$ and $\Delta\Gamma_{s}$ measurements shown in Fig.~\ref{Fig:MainPlot} is found to be $\phi_{s}=-21\pm31$~mrad and $\Delta\Gamma_{s}=0.085\pm0.006$~ps$^{-1}$~\cite{bib:MainPlot}. It is dominated by the measurements from the LHCb collaboration and is consistent with the SM predictions. However, the combined measurement is still far from the SM precision thus leaving some room for NP effects. The improvements on the sensitivity of $\phi_{s}$ are expected from the inclusion of data collected in 2015-2018 at center-of-mass energy of $\sqrt{s}$~=~13~TeV. It will allow to use the $b\to c$ and $b\to s$ processes with very small branching fraction to constraint the $\phi_{s}$ measurement.

\section{Future contributions for measuring $\phi_{s}$ and $\Delta\Gamma_{s}$}
\label{sec-2}
\subsection{Observation of the $B^{0}_{s}\to\eta_{c}\phi$ decay}

\begin{figure}[htb]
\begin{minipage}{0.48\linewidth}
  \centerline{\includegraphics[width=120pt]{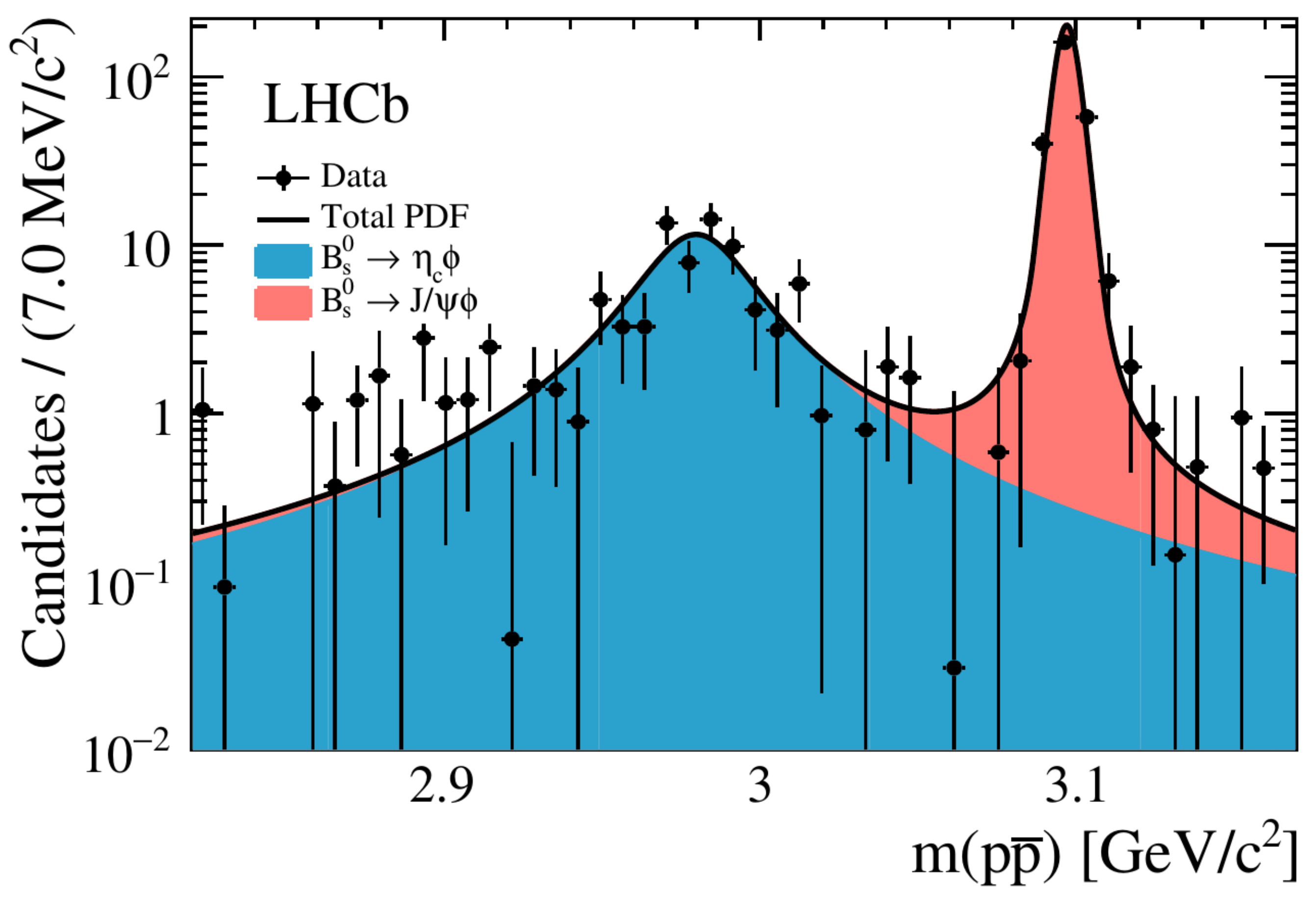}}
 \end{minipage}
 \hfill
   \begin{minipage}{0.48\linewidth}
  \centerline{\includegraphics[width=120pt]{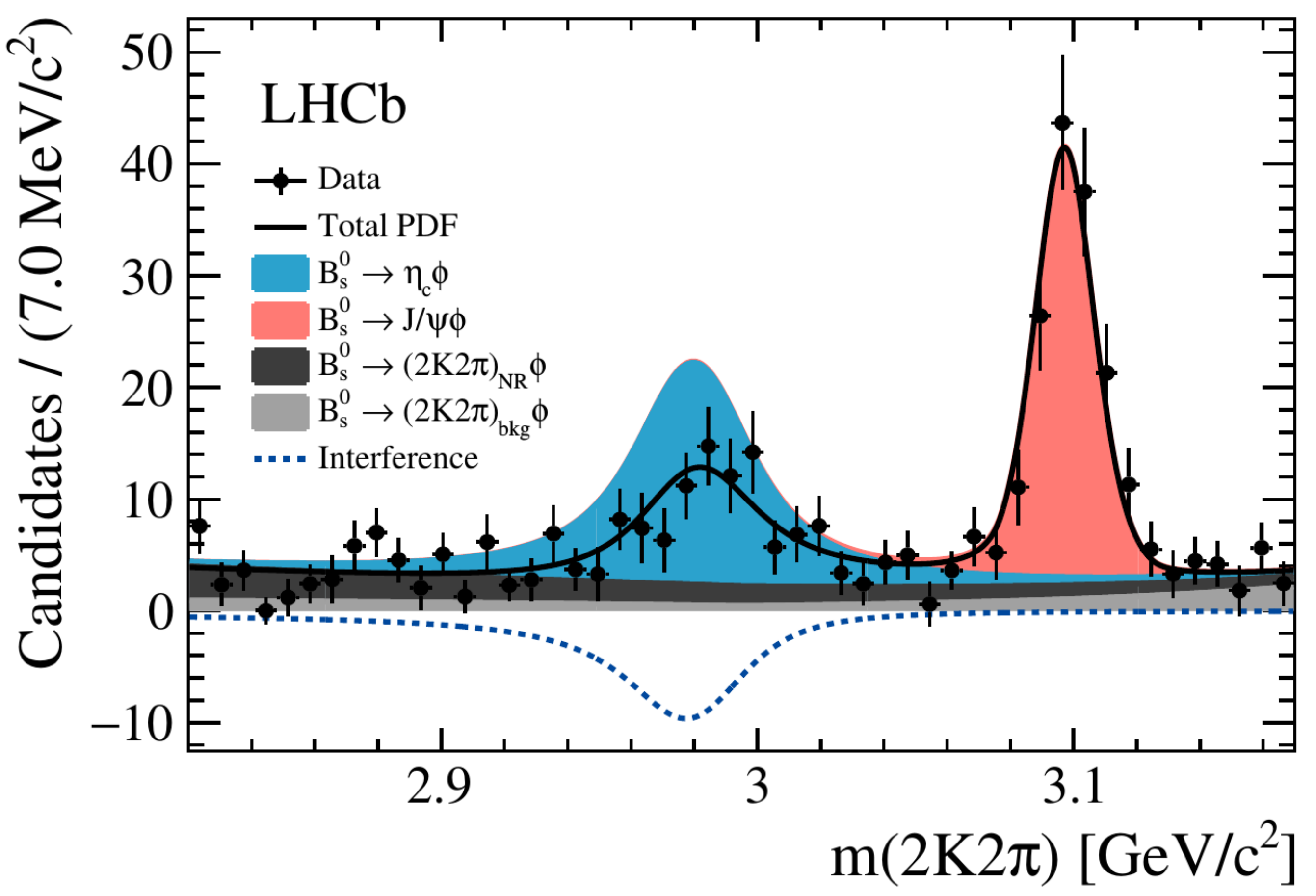}}
 \end{minipage}
 \vfill
 \begin{minipage}{0.48\linewidth}
  \centerline{\includegraphics[width=120pt]{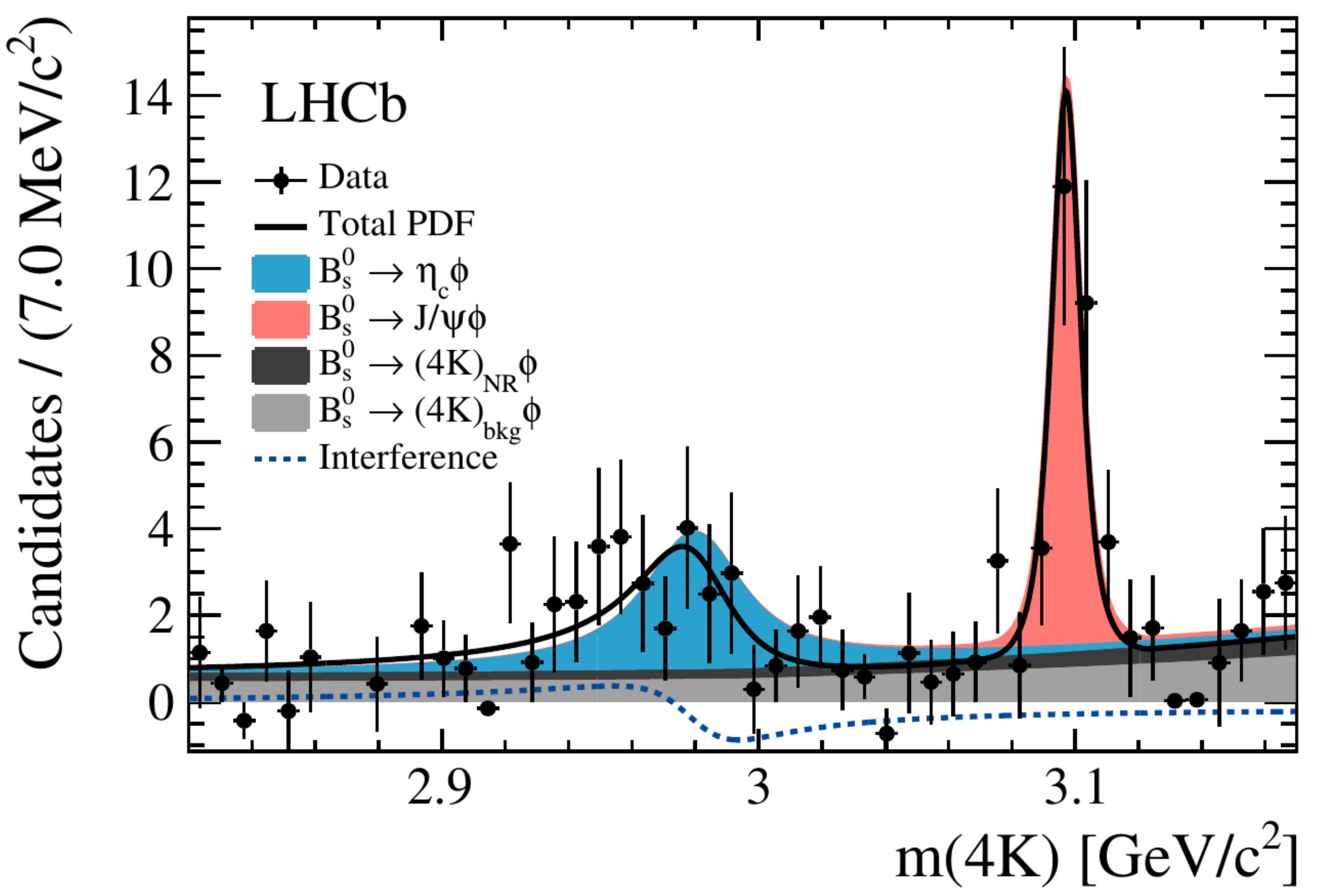}}
 \end{minipage}
 \hfill
   \begin{minipage}{0.48\linewidth}
  \centerline{\includegraphics[width=120pt]{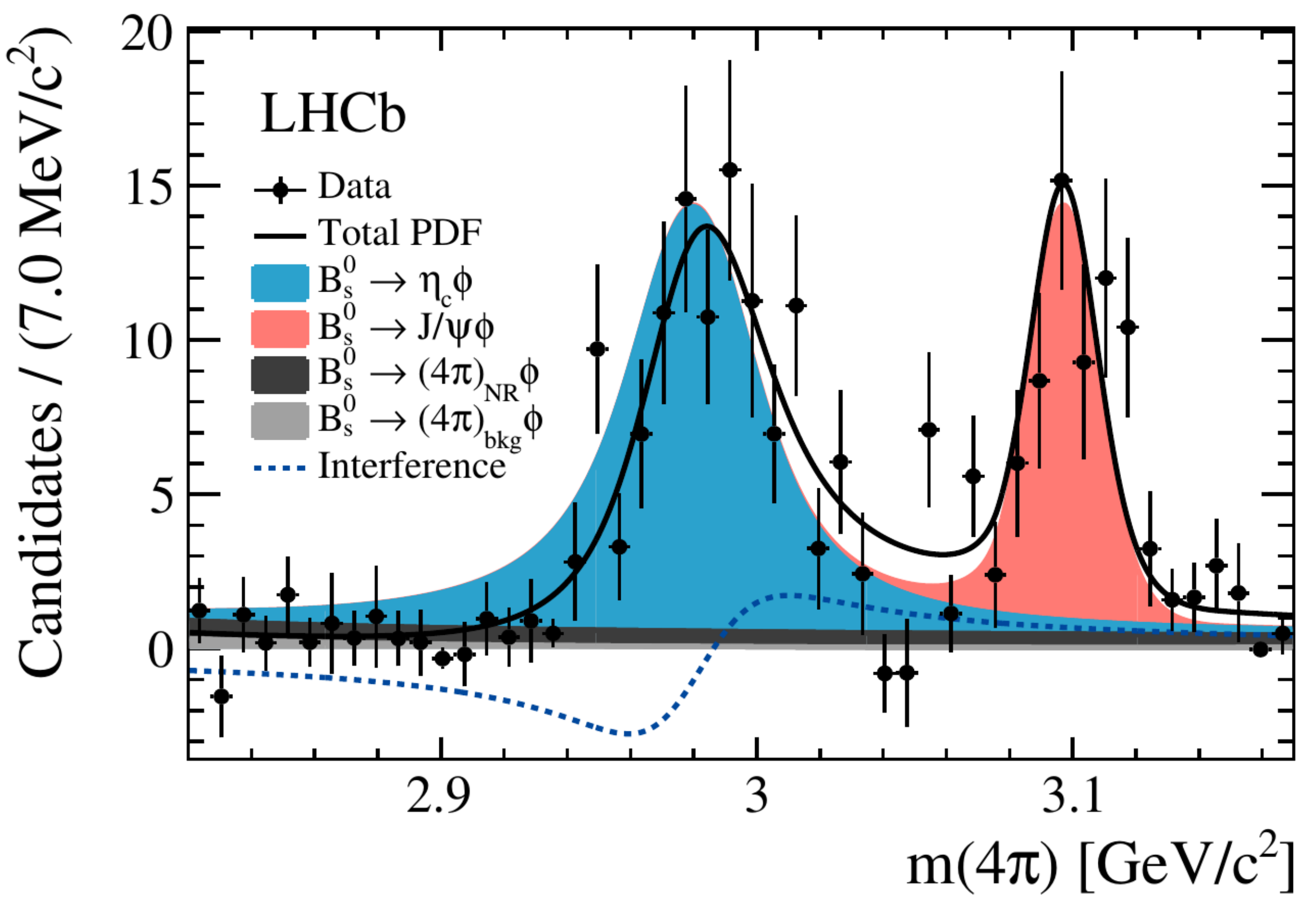}}
 \end{minipage}
\caption{Invariant mass distributions for selected $p\bar{p}$, $K^{+}K^{-}\pi^{+}\pi^{-}$, $K^{+}K^{-}K^{+}K^{-}$ and $\pi^{+}\pi^{-}\pi^{+}\pi^{-}$ combinations.}
\label{Fig:Etac}
\end{figure}

The $B^{0}_{s}\to\eta_{c}\phi(\to K^{+}K^{-})$ decay mode, with $\eta_{c}\to K^{+}K^{-}\pi^{+}\pi^{-}$, $K^{+}K^{-}K^{+}K^{-}$, $\pi^{+}\pi^{-}\pi^{+}\pi^{-}$ and $p\bar{p}$, has been observed by the LHCb collaboration for the first time~\cite{Aaij:2017hfc}. This decay is an another $\bar{b}\to \bar{c}c\bar{s}$ transition that could be used to measure $\phi_{s}$. The interference between the $\eta_{c}$ and purely nonresonant contributions is taken into account using an amplitude model to simultaneously fit the four hadrons and $p\bar{p}$ mass distributions (Fig.~\ref{Fig:Etac}). The branching fraction is normalized to the $J/\psi$ mode and found to be $\mathcal{B}(B^{0}_{s}\to \eta_{c}\phi)=[5.01\pm0.53(\mathrm{stat})\pm0.27(\mathrm{syst})\pm0.63(\mathcal{B})]\times10^{-4}$. First evidence for the $B^{0}_{s}\to\eta_{c}\pi^{+}\pi^{-}$ decay mode has also been reported, with a branching fraction of $\mathcal{B}(B^{0}_{s}\to\eta_{c}\pi^{+}\pi^{-})=[1.76\pm0.59(\mathrm{stat})\pm0.12(\mathrm{syst})\pm0.29(\mathcal{B})]\times10^{-4}$.

\subsection{$B^{0}_{s}\to J/\psi\eta$ effective lifetime}

\begin{figure}[htb]
\begin{minipage}{0.48\linewidth}
  \centerline{\includegraphics[width=140pt]{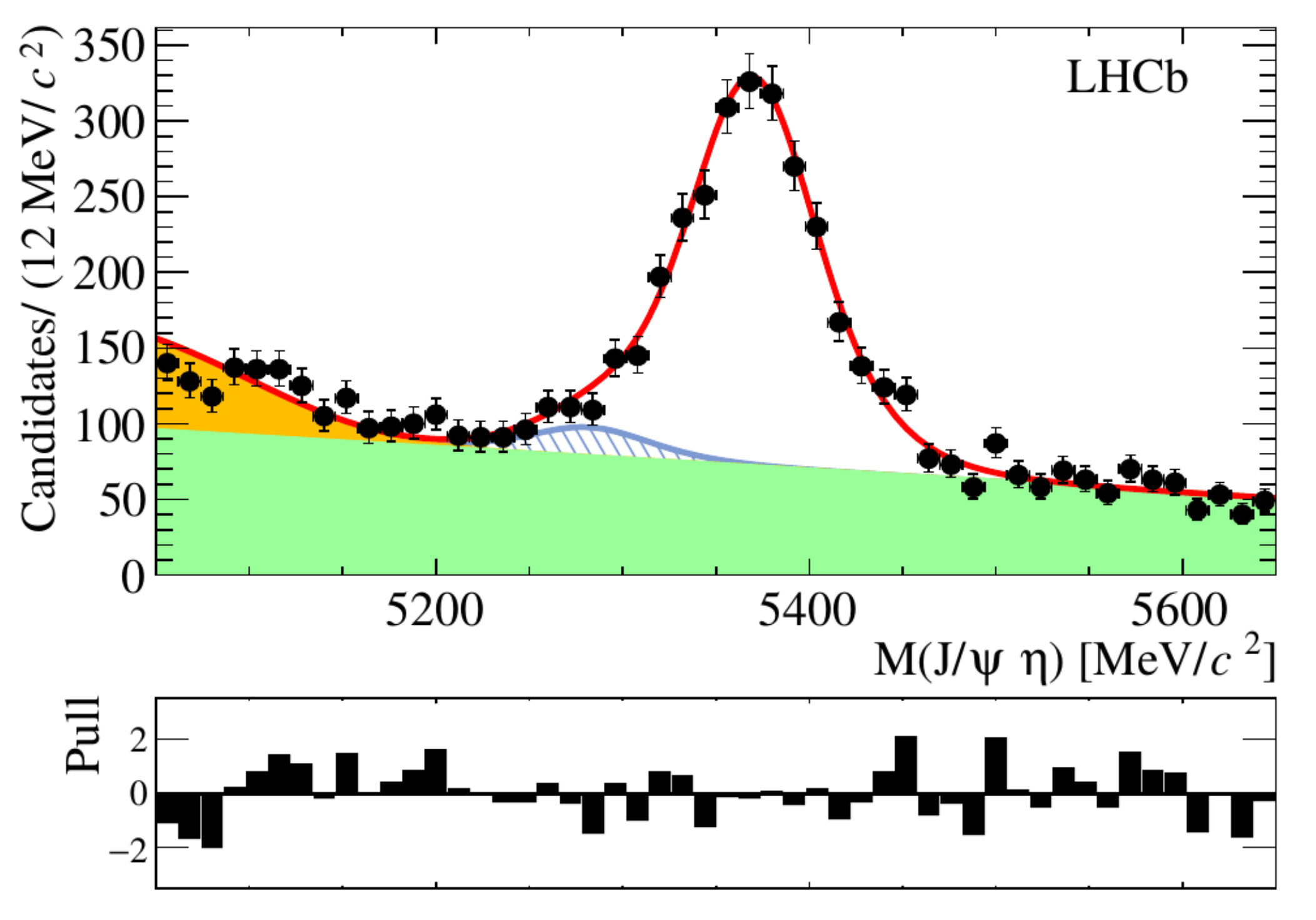}}
 \end{minipage}
 \hfill
   \begin{minipage}{0.48\linewidth}
  \centerline{\includegraphics[width=140pt]{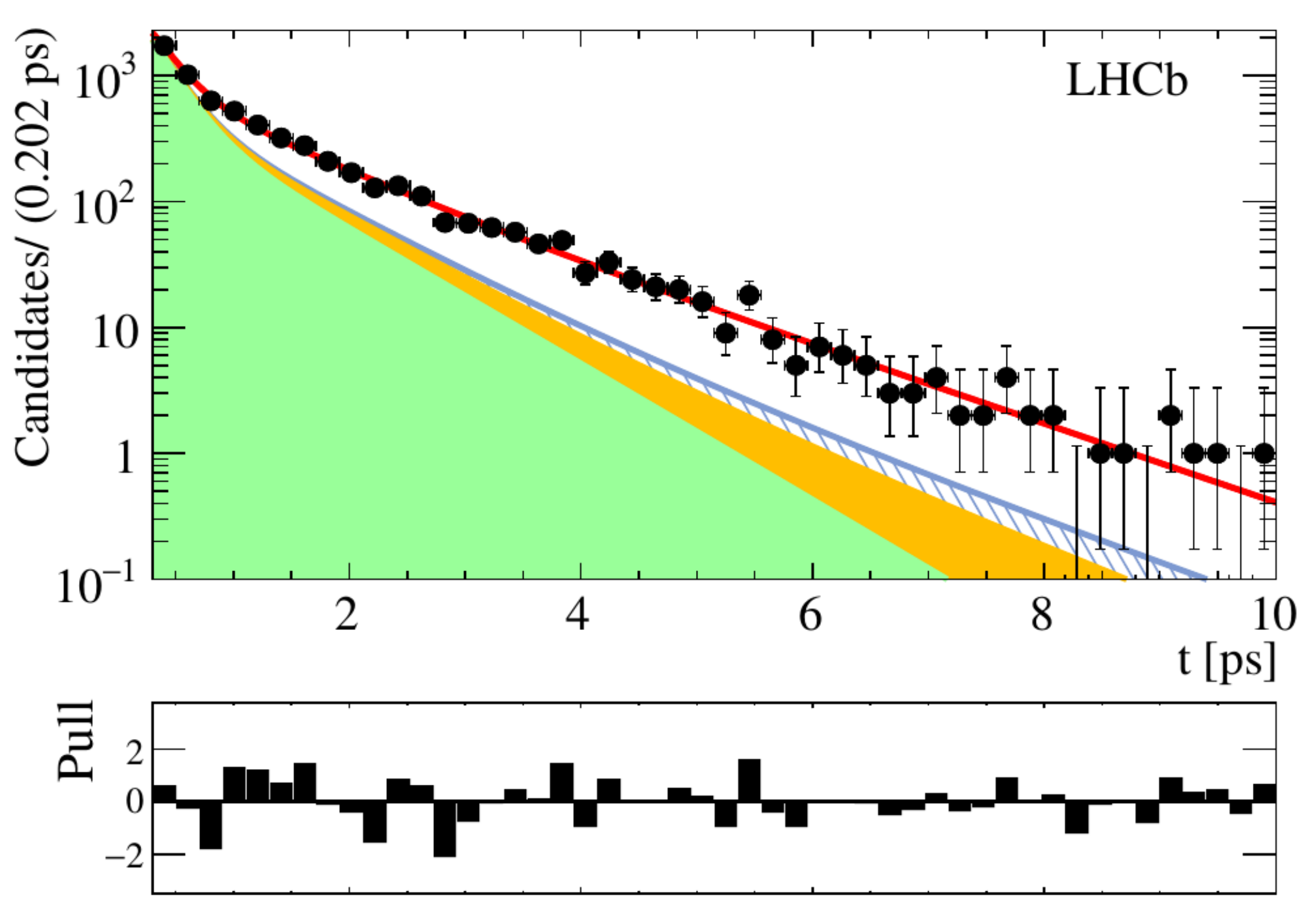}}
 \end{minipage}
\caption{Distributions of $J/\psi\eta$ invariant mass (left) and decay time (right) for selected $B^{0}_{s}\to J/\psi\eta$ decays. Combinatorial background (green), background from $B^{0}\to J/\psi\eta$ decays (blue) and partially reconstructed background (orange) are shown.}
\label{Fig:JpsiEta}
\end{figure}

The $B^{0}_{s}$ effective lifetime has been measured by the LHCb collaboration using the $C\!P$-even $B^{0}_{s}\to J/\psi(\to\mu^{+}\mu^{-})\eta(\to\gamma\gamma)$ decay mode using Run1 data~\cite{Aaij:2016dzn}. As $\phi_{s}$ is measured to be small and assuming $C\!P$ conservation, the effective lifetime corresponds to the lifetime of the light $B^{0}_{s}$ mass eigenstate, $\tau_{L}\propto\Gamma_{L}$. The invariant mass resolution is approximately 48~MeV/c$^{2}$ (Fig.~\ref{Fig:JpsiEta}) causing the overlap of the $B^{0}_{s}$ signal mode with the $B^{0}\to J/\psi\eta$ background component. The effective lifetime for $\sim$3000 signal candidates is measured to be $\tau_{\mathrm{eff}}=1.479\pm0.034\pm0.011$~ps. The result is consistent with, and has a similar precision to, the other $C\!P$-even lifetime measurements~\cite{Aaij:2013bvd,Aaij:2014fia}.     

\subsection{Observation of the $B^{0}_{s}\to\phi\pi^{+}\pi^{-}$ decays}

\begin{figure}[htb]
\centering
\begin{minipage}{0.48\linewidth}
  \centerline{\includegraphics[width=150pt]{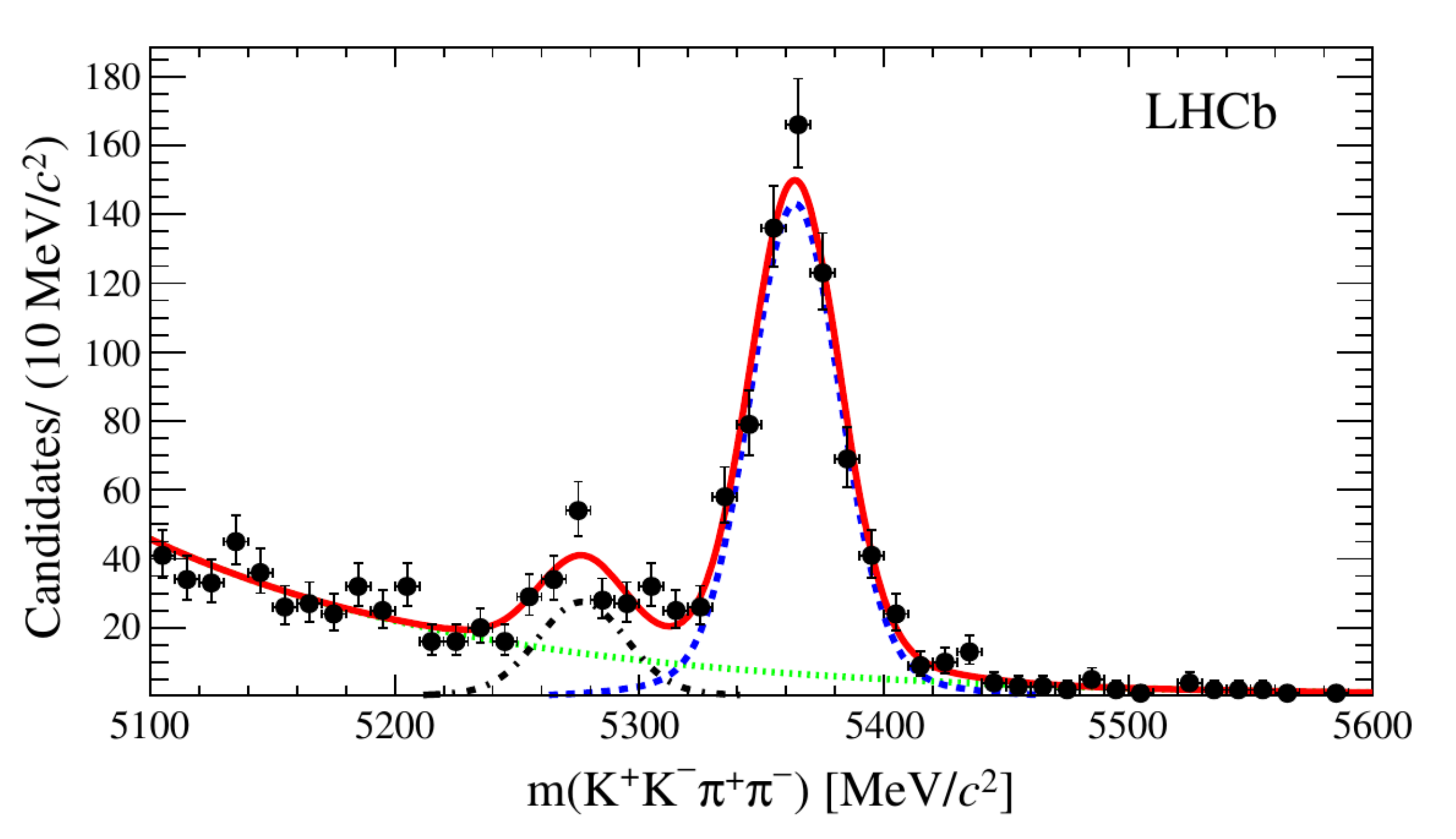}}
 \end{minipage}
 \hfill
   \begin{minipage}{0.48\linewidth}
  \centerline{\includegraphics[width=130pt]{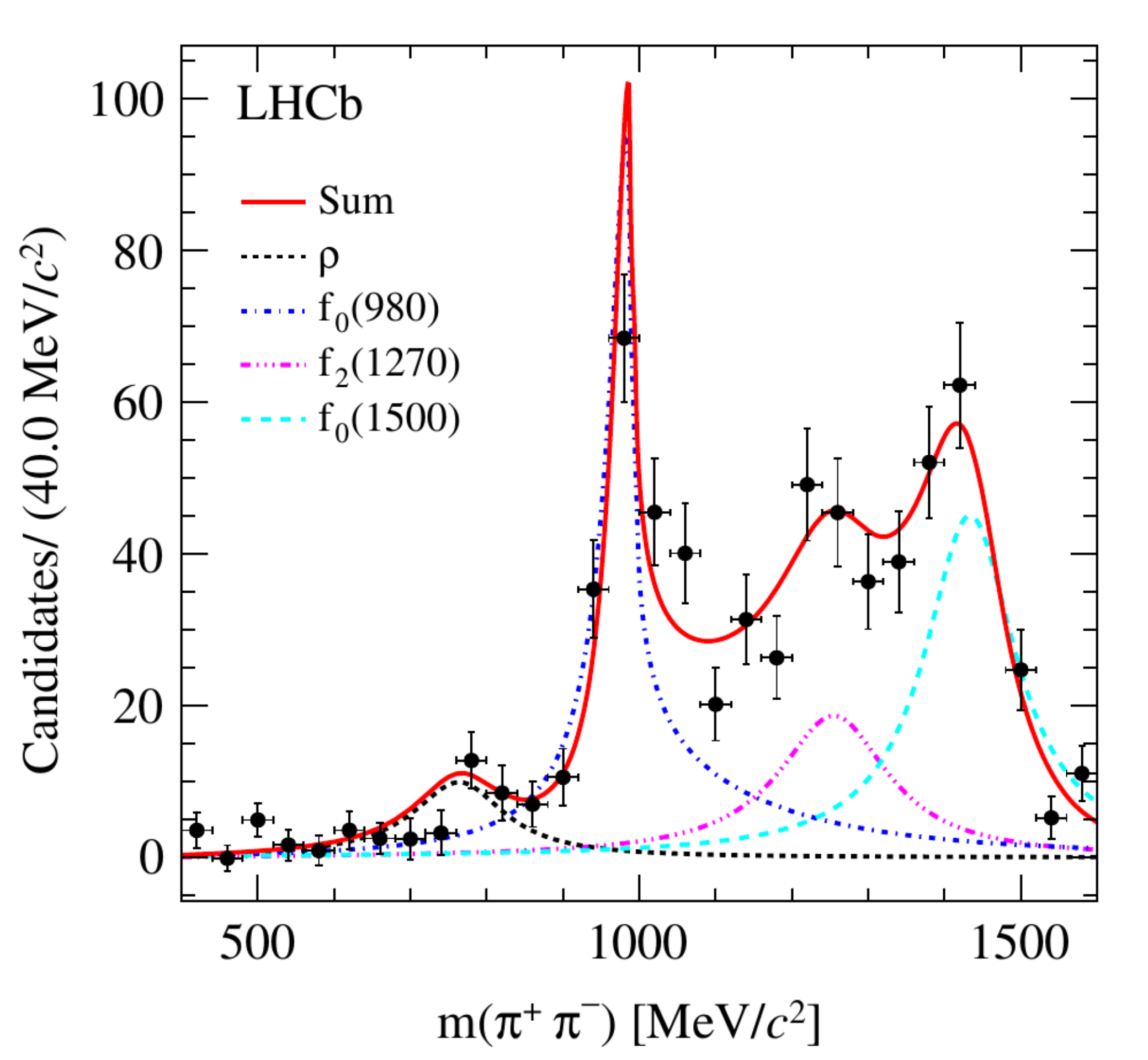}}
 \end{minipage}
\caption{(left) Distribution of $K^{+}K^{-}\pi^{+}\pi^{-}$ invariant mass where the (blue) dashed line is the $B^{0}_{s}$ signal, the (green) dotted line shows the combinatorial background and the (black) dot-dashed line indicates the $B^{0}$ component. (right) Distributions of $\pi^{+}\pi^{-}$ invariant mass with contributing components.}
\label{Fig:PhiPiPi}
\end{figure}

The first observation of the inclusive decay $B^{0}_{s}\to\phi(\to K^{+}K^{-})\pi^{+}\pi^{-}$ has been performed by the LHCb collaboration~\cite{Aaij:2016qnm}. Fig.~\ref{Fig:PhiPiPi} shows the result of the final fit to the $m(K^{+}K^{-}\pi^{+}\pi^{-})$ distribution. The $B^{0}_{s}$ yield is found to be around 700 signal candidates. Since the $\pi^{+}\pi^{-}$ spectrum includes several resonances, an amplitude analysis to the $\pi^{+}\pi^{-}$ mass and decay angle distributions is used to separate out exclusive contributions to the $B^{0}_{s}$ meson decays (Fig.~\ref{Fig:PhiPiPi}). The $B^{0}_{s}\to\phi\phi$ is used as a normalization channel for both the inclusive and exclusive decays. The measurement of their branching fractions is $\mathcal{B}(B^{0}_{s}\to\phi f_{0}(980))=[1.12\pm0.16^{+0.09}_{-0.08}\pm0.11]\times10^{-6}$, $\mathcal{B}(B^{0}_{s}\to\phi f_{2}(1270))=[0.61\pm0.13^{+0.12}_{-0.05}\pm0.06]\times10^{-6}$ and $\mathcal{B}(B^{0}_{s}\to\phi\rho^{0})=[2.7\pm0.7\pm0.2\pm0.2]\times10^{-7}$, where the first uncertainty is statistical, the second is systematic, and the third is related to the knowledge of the normalization channel branching fraction. The decays $\phi f_{0}(980)$, $\phi f_{2}(1270)$ and $\phi\rho^{0}$ are observed with a significance of 8$\sigma$, 5$\sigma$ and 4$\sigma$, respectively. The measurements are consistent with the SM predictions and, in case of the $B^{0}_{s}\to\phi\rho^{0}$, they provide a constraint on possible contributions from NP effects~\cite{Hofer:2010ee}.

\subsection{$C\!P$ asymmetry measurement of the $B^{\pm}\to J/\psi\rho^{\pm}$ decays}

\begin{figure}[htb]
\centering
\begin{minipage}{0.48\linewidth}
  \centerline{\includegraphics[width=140pt]{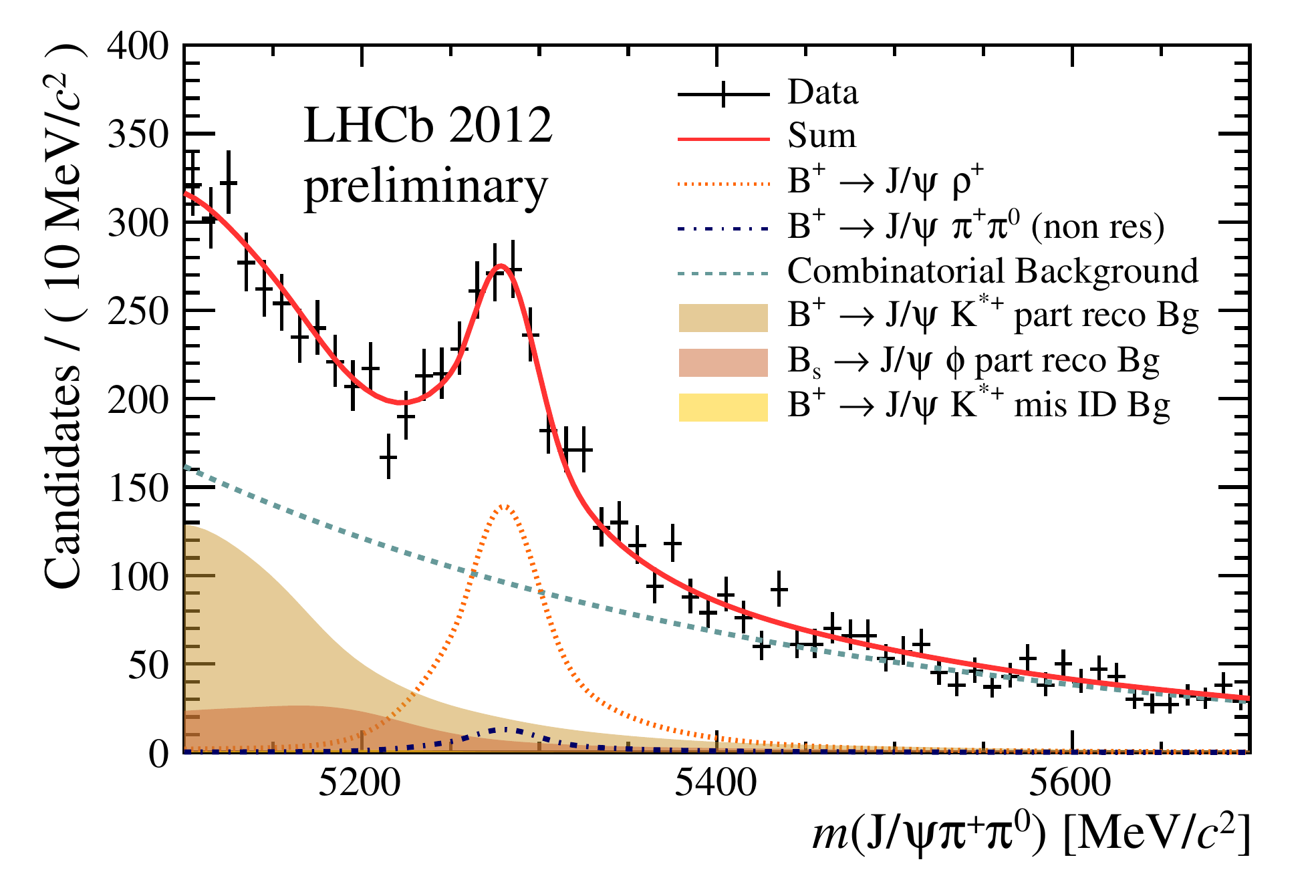}}
 \end{minipage}
 \hfill
   \begin{minipage}{0.48\linewidth}
  \centerline{\includegraphics[width=140pt]{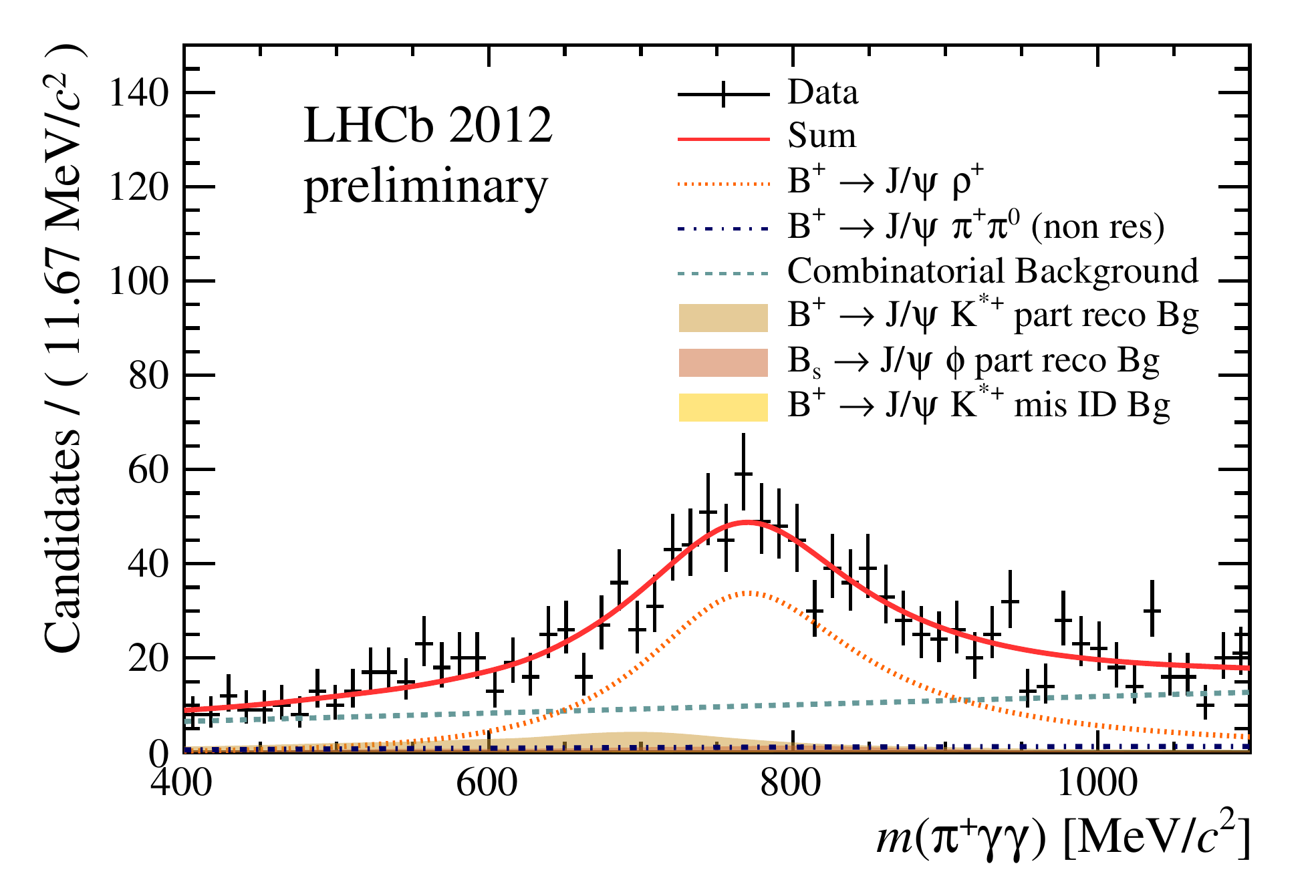}}
 \end{minipage}
\caption{(left) Distribution of $J/\psi\pi^{+}\pi^{0}$ invariant mass for 2012 data set. (right) Distribution of $\pi^{+}\gamma\gamma$ invariant mass for 2012 data set for $m_{J/\psi\pi^{+}\pi^{0}}\in(5250,5310)$~MeV/c$^{2}$ range.}
\label{Fig:JpsiRho}
\end{figure}

The branching fraction and direct $C\!P$ asymmetry of the $B^{\pm}\to J/\psi(\to\mu^{+}\mu^{-})\rho^{\pm}(\pi^{\pm}\pi^{0})$ decay have been measured by the LHCb collaboration~\cite{Aaij:2018abc}. The decay predominantly proceeds via a $b\to c\bar{c}d$ transition involving tree and penguin amplitudes. The measurement of $\mathcal{A}^{C\!P}$ provides an estimate of imaginary part of the penguin-to-tree amplitude ratio of the $b\to c\bar{c}d$ transition, which can be used to place constraints on penguin effects in measurements of the $C\!P$-violating phase $\phi_{s}$, assuming SU(3) symmetry. The $B^{\pm}\to J/\psi K^{\pm}$ decay is used as a normalisation channel for the branching fraction measurement. The fit to $B^{+}$ and $\rho^{+}$ candidate mass has been performed to distinguish $J/\psi\rho^{+}$ from non-resonant $J/\psi\pi^{+}\pi^{0}$ decay as shown in Fig.~\ref{Fig:JpsiRho}. The branching fraction and $C\!P$ asymmetry are measured to be $\mathcal{B}(B^{+}\to J/\psi\rho^{+})=(3.79^{+0.25}_{-0.24}\pm0.32)\times10^{-5}$ and $\mathcal{A}^{C\!P}(B^{+}\to J/\psi\rho^{+})=-0.045^{+0.056}_{-0.057}\pm0.008$. The results are consistent with BaBar measurement~\cite{Aubert:2007xw}. The measured value of $C\!P$ asymmetry is consistent with the corresponding measurement using $B^{0}\to J/\psi\rho^{0}$ decays, as expected from isospin symmetry~\cite{Aaij:2014vda}.

\section{Summary}
\label{sec-3}

The most precise measurement of $C\!P$-violating phase $\phi_{s}$ and decay width difference $\Delta\Gamma_{s}$ in the $B^{0}_{s}$ system has been performed using Run1 data collected by LHCb experiment corresponding to an integrated luminosity of 3~fb$^{-1}$. So far all results are compatible with the SM prediction. In order to reach an uncertainty of the measurement comparable or even better than the theoretical uncertainty of the SM prediction aside from improvements in available luminosity, inclusion of new decay modes has been explored. For example, the $B^{0}_{s}\rightarrow J/\psi(\to e^{+}e^{-})\phi$ channel not only could bring about 10$\%$ of the $\mu^{+}\mu^{-}$ mode statistics, but it will be also an important verification of the $B^{0}_{s}\rightarrow J/\psi(\to \mu^{+}\mu^{-})\phi$ as kinematics for both channels are expected to be identical. The statistical sensitivity to $\phi_{s}$ after the LHCb upgrade, with an integrated luminosity of 50~fb$^{-1}$, is expected to be $\sim$9~mrad. As that will be close to the present theoretical uncertainty~\cite{bib:Upgrade}. As the measurement precision improves, the penguin pollution contributions to the $B^{0}_{s}$ meson decays need to be kept under control. Current measurements constrain them to be smaller than 20~mrad~\cite{Aaij:2015mea,Aaij:2014vda,Aaij:2015tza}.

\begin{center}
{\bf Acknowledgments}
\end{center}

The talk has been supported by the MNiSW with grant DIR/WK/2016/16. I would like to thanks the organizers of the CKM2018 workshop for the invitation to present this work and my LHCb colleagues who helped in the preparation of this talk.

\bibliography{bib-list}

\end{document}